\documentclass[11pt, a4paper, twoside, openright]{book} %draft

\usepackage{graphicx,color}
\usepackage{amssymb, amsmath, array, times}
\usepackage{url}
\usepackage{amsfonts}
\usepackage{graphicx,epstopdf}

\newtheorem{definition}{Definition}

\begin{document}
\frontmatter
% Example of title page for the projects carried out within the lasec 

% Simply include it in your mastex tex file: 
%        \input{cover}

% Updated March 2006 (SP)

\newcommand{\logoepfl}[0]{
  \begin{center}
    \includegraphics[width=4cm]{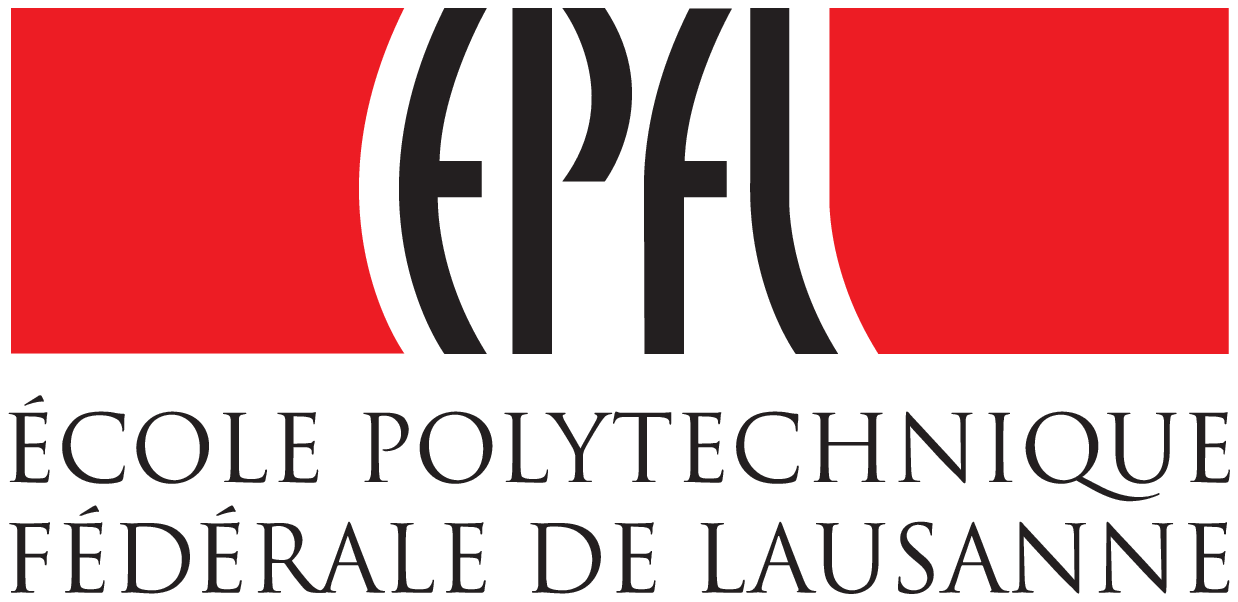}
  \end{center}
  \vspace{0.3cm}
  \hrule
}
\newcommand{\logolasec}[0]{
  \vspace{1cm}
  \hrule
  \begin{center}
    \includegraphics[width=4.5cm]{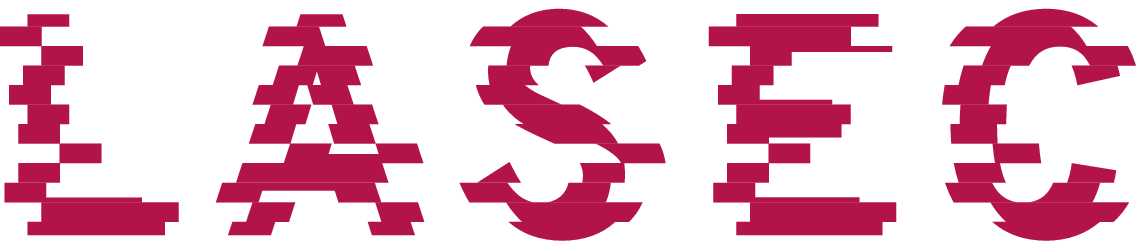}
  \end{center}
}
\newcommand{\project}[1]{
  \begin{center}
    \large{#1}
  \end{center}
  \vspace{1cm}
}
\newcommand{\department}[1]{
  \begin{center}
    \large{#1}
  \end{center}
}
\newcommand{\supervisor}[3]{
  \begin{center}
    \begin{normalsize}{
        \bf #1}\\#2\\#3
    \end{normalsize}
  \end{center}
}
\renewcommand{\author}[1]{
  \begin{center}
    \Large{#1}
  \end{center}
  \vspace{0.5cm}
}
\renewcommand{\title}[1]{
  \vspace{3cm}
  \begin{center}
    \huge{#1}
  \end{center}
  \vspace{1.7cm}
}
\renewcommand{\date}[2]{
  \begin{center}
    \normalsize{#1 #2}
  \end{center}
  \vspace{0.5cm}
}

\thispagestyle{empty}

% begin title page
  \logoepfl
  
  \title{Selling train tickets by SMS}
  
  \author{Steven Meyer}
  \department{School of Computer and Communication Sciences}
  \project{Semester Project}
  
  \date{June}{2010}

  \begin{center}
    \begin{tabular}{cc}
      \begin{tabular}{p{4.0cm}}
        \supervisor{Responsible}{Prof. Serge Vaudenay}{EPFL / LASEC}
      \end{tabular}&
      \begin{tabular}{p{4.0cm}}
        \supervisor{Supervisor}{Khaled Ouafi}{EPFL / LASEC}
      \end{tabular}
    \end{tabular}
  \end{center}

\logolasec
% end title page

\cleardoublepage
\chapter*{Abstract} 

Selling train tickets has evolved in the last ten years from queuing in the railway station, to buying tickets on the internet and printing them. Both alternatives are still viable options, though they are time consuming or need printing devices. Nowadays it is essential to offer a service that is as fast and efficient as possible: mobile phones provide an accessible, affordable and widely available tool for supplying information and transferring data.

The goal of this project is to design a train ticket contained in a SMS message. While there are several challenges related to the project, the main one is the security and how we can digitally sign a train ticket that is contained in 160 characters. The solution offered in this project is the implementation of the MOVA Signature (from the name of the inventors MOnnerat and VAudenay) that uses an interactive verification and therefore allows signature of 20 bits (roughly 4 characters).

\tableofcontents
\mainmatter
\chapter{Introduction}
 In 1976, when Diffie and Hellman published their paper on public cryptosystem \cite{hellman1976new}, they explained the concept of digital signature that corresponds for the electronic world to the handwritten signature. To implement a digital signature we use asymmetric cryptography with one private key and one public key. The private key used to decrypt or sign a message is different from the public one used to encrypt or verify, though both are related through a mathematical function; additionally, finding the private key on the basis of the public one must require complex calculations that cannot be realized in a reasonable time. In general the Signer uses his private key to sign the object and then anyone can use the public key to verify the validity of the signature. This system offers authentication by guaranteeing that the sender is the person he pretends to be and integrity by guaranteeing that the receiver gets what the sender has sent.\

In 1989, Chaum and van Antwerpen proposed Undeniable Signatures \cite{chaum1989undeniable} as a system that would not be universally verifiable (verifiable just by having the public key), but that would necessitate an interactive protocol between the Signer and the Verifier to be able to confirm or deny a signature. \

In 2004,  Monnerat and Vaudenay proposed an undeniable signature scheme named MOVA \cite{monnerat2004generic} that has the particularity to generate very short signatures while still maintaining a high security.\

Switzerland is well known for its efficient railway service which is reliable and punctual though nor very fast. Tickets bought on board are taxed with a penalty fine, which leaves passengers with the options of queuing at the station, buying the tickets through internet and printing them, or receiving the tickets by MMS. \

MMS is a protocol that allows to send/receive rich media up to several kilobytes on a cell phone and extends the SMS system which is limited to 140 bytes. The MMS service uses the data connection of the cellphone and therefore works only on modern phones with operators that support this protocol and of course is much more expensive (especially while roaming) than SMS.

\chapter{Preliminaries}
\section{Mathematical Background}
To understand the rest of the paper, we will first have to define some mathematical concepts:

\subsection{Algorithmic Complexity}
\begin{definition}[Negligible]
A function $f : \mathbb{N} \rightarrow \mathbb{R}^+$ is called \textit{negligible} ($negl(\cdot)$) if for any polynomial $p : \mathbb{N} \rightarrow \mathbb{R}^+$ there exists an integer $n_0$ such that $n \geq n_0 \Rightarrow f(n)< \frac{1}{p(n)} $.
\end{definition}

\begin{definition}[Efficient] 
An algorithm is said to be efficient if its running time is bounded by a polynomial of the size of its inputs.
Conversely, an operation is said to be efficient if it is carried by an efficient algorithm.
\end{definition}

\subsection{Groups}

A group $(G,\ast)$ is an algebraic structure defined as a set $G$ together with a binary operation $\ast$: $G \ast G \rightarrow G$. We write ``$a \ast b$" for the result of applying the operation $\ast$ on the two elements $a$ and $b$ of $G$. To have a group, $\ast$ must satisfy the following axioms:

\begin{itemize}
\item \textbf{Closure} $ \forall a,b \in G: a \ast b \in G$.
\item \textbf{Associativity} $\forall a,b,c \in G: (a \ast b) \ast c =a \ast (b \ast c)$.
\item \textbf{Identity element} $ \exists e \in G : \forall a \in G, e \ast a = a = a \ast e$. 
\item \textbf{Inverse element} $ \forall a \in G, \exists b \in G : a \ast b = e = b \ast a$ (where $e$ is the identity element).
\end{itemize}

\subsection{Group Homomorphism}
Given two groups $(G, \ast)$ and $(H, \cdot)$, a group homomorphism from $(G, \ast)$ to $(H, \cdot)$ is a function $h$ : $G \rightarrow H $ such that $ \forall a,b \in G $ : $h(a \ast b) = h(a) \cdot h(b)$. 

\begin{definition}
[Interpolation] Given two groups $(G, \ast)$ and $(H, \cdot)$, and a subset  $S:=\{(x_1,y_1) \ldots (x_s,y_s)\} \subseteq G \times H $, we say that the set of points $S$ \textit{interpolates} in a group homomorphism if there exists a group homomorphism $h$ : $G \rightarrow H $ such that $ h(x_i)=y_i,  \forall i=1 , \ldots , s $.

Given $A \subseteq G \times H$ and $B \subseteq G \times H$, we say that $A$ interpolates in a group homomorphism with $B$ if $A \cup B$ interpolates in a group homomorphism.
\end{definition}

\subsubsection{n-S-Group Homomorphism Interpolation Problem}
n-S-GHI problems uses two groups $G$ and $H$, a set of points $S \subseteq G \times H$ and a positive integer $n$. Given $x_1, \ldots ,x_n$ chosen randomly, the Problem is to compute $y_1, \ldots ,y_n \in H $ such that $ {(x_1, y_1), \ldots , (x_n, y_n)}$ interpolates with $S$ in a group homomorphism.

\subsubsection{n-S-Group Homomorphism Interpolation Decisional Problem}
n-S-GHID problem uses two groups $G$ and $H$, a set of points $S \subseteq G \times H$ and a positive integer $n$. The problem is to \textit{decide} if an instance $I$ is generated by a set of points ${(x_1, y_1),\ldots , (x_n, y_n)} \in X \times Y$ chosen at random such that it interpolates with $S$ in a group homomorphism or is generated by randomly choosing $n$ couples in $(G \times H)^n$.

\begin{definition}[H-generation of G]
Let two groups $G$ and $H$ with  $x_1,\ldots ,x_n \in G$ and $y_1, \ldots , y_n \in H$, we say that $x_1,\ldots ,x_n$ H-generate of G if there exists  \textit{at most one}  homomorphic function $h $ such that 
$$\forall i = 1, \ldots , n, h(x_i)=y_i $$
\end{definition}

\begin{definition}[Expert Group Knowledge]
Let $G$ and $H$ be two groups, let $d$ be the cardinality of $H$, and the set $S=\{x_1, \ldots , x_n\} \in G^n$, such that they H-generate G. We say an algorithm has an \textit{Expert Group Knowledge} of $G$ with the set $S$ if it is able to find for a random $ x \in G$ coefficients $a_1 , \ldots ,a_n \in \mathbb{Z}_d$ and $r \in G$ such that 
$$x = dr + a_1x_1 + \ldots + a_sx_s$$
\end{definition}

\section{Cryptographic notions}
We will now see some Cryptographic definitions and properties that will be used later on in the paper

\subsection{Statistical Distances and Distinguishers}

\begin{definition}[Statistical Distance]
The statistical distance $\Delta$ between two discrete random variables
$X_1$ and $X_2$ is defined as: 
$$\Delta(X_1,X_2):=\frac{1}{2} \sum_x(|\Pr[X_1=x]-\Pr[X_2=x]|)$$
\end{definition}

\begin{definition}[Distinguisher]
Given two distributions $X_0$ and $X_1$, a distinguisher, denoted $\mathcal{D}$ hereafter, is an algorithm given either $X_0$ or $X_1$, chosen randomly, which tries to guess which of the two is given.

Concretely, we consider the following success probability
$$
|\Pr [\mathcal{D}(1^k,X_0)=1]-\Pr[\mathcal{D}(1^k,X_1)=1]|,
$$
for which we consider several cases:
\begin{enumerate}
\item if the probability is equal to $0$, we say that $X_0$ and $X_1$ are perfectly indistinguishable.
\item if the probability is lower than some value $\epsilon$, we say that $X_0$ and $X_1$ are $\epsilon$-statistically indistinguishable.
\item if the probability is negligible in $k$ and $\mathcal{D}$ is polynomially bounded in $k$, we say that $X_0$ and $X_1$ are computationally indistinguishable.
\end{enumerate}
\end{definition}

\subsection{Pseudorandom Generator}
Pseudorandom generator defines a deterministic procedure that takes as parameter a seed $s$ (that can be random) and produces an output sequence that is indistinguishable from a truly random sequence.

\subsection{One-way Functions}

\begin{definition}[One-way and Trapdoor-one-way functions] 
A function $f:G \rightarrow H$ is said to be one-way if it is efficiently computable, i.e., given $x\in G$, computing $y=f(x)$ is efficient.
However, inverting the function is ``hard'', in the sense that

$$
\forall \mathcal{A}, \Pr[a \leftarrow \mathcal{A}(1^k,b) | b=f(a)]= negl(k)
$$
where $\mathcal{A}$ is polynomially bounded and the probability is taken over the randomness of $\mathcal{A}$ and the random choice of $a$.

If there exists some secret information $s$ for which the invertibility of $f$ becomes efficient, we say that $f$ is a trapdoor-one-way function.
\end{definition}

\subsection{Hash Function}

A hash function is an efficiently calculable function $h: \{0;1\}^* \rightarrow \{0;1\}^h$ that reduces a message of an arbitrary length to a given length that has these proprieties:

\begin{itemize}
  \item Preimage Resistance: given $y \in \{0;1\}^h$ it is difficult to calculate $x$ such as $f(x)=y$.
  \item Second Preimage Resistance: given $x_{1}$ it is difficult to find $x_{2}$ (with $x_{1} \neq x_{2}$) such as $f(x_{1})=f(x_{2})$.
  \item Collision Resistance: it is difficult to find $x_1, x_2$ such that $f(x_1)=f(x_2)$.
\end{itemize}

\subsection{Commitment Scheme}
A Commitment Scheme is a protocol in which one generates a commitment value from a data that binds him to not change it but that does not reveal information about the data. The commitment then can later be opened to reveal the data. A Commitment Scheme is defined by: a message (data) $m \in M$ in the space of messages; a commitment $c \in C$ in the space of Commitments; a decommitment $d \in D$ in the space of Decommitment (that can reveal the committed data); a committing algorithm $\mathsf{Commit}$ and an opening algorithm $\mathsf{Open}$ that are used below:
\begin{itemize}
  \item $\mathsf{Commit}(m) \rightarrow (c,d)$ the probabilistic commitment algorithm generates from the message $m$ a commitment $c$ and a decommitment $d$.
  \item $\mathsf{Open}(m,c,d) \rightarrow (\mathsf{true}$ or $\mathsf{false})$ the usually deterministic $\mathsf{Open}$ algorithm verifies  if $m$ is a valid message for the given commitment $c$ and decommitment $d$.
\end{itemize}
A commitment Scheme should have the following:
\begin{itemize}
  \item Completeness: if the Commiter and the Opener behave as specified, the opener always accepts the proof:
  $$\Pr [\mathsf{Open}(m,c,d) \rightarrow \mathsf{true} | \mathsf{Commit}(m) \rightarrow \mathsf{true}]=1$$

 \item Binding Property: one cannot produce two distinct messages $m, m'$, two decommitment values $d, d'$, one commitment value $c$ such that $ \mathsf{Open}(m,c,d) \rightarrow \mathsf{true}$ and $\mathsf{Open}(m',c,d') \rightarrow \mathsf{true}$.
 
 Concretely, we consider the following success probability
 $$\Pr [\mathsf{Open}(m',c,d') \rightarrow \mathsf{true}|(c,d) \leftarrow \mathsf{Commit}(m), m' \neq m]$$
 for which we consider several cases:
 \begin{enumerate}
 \item if the probability is equal to $\frac{1}{|M|}$ we say that the commitment function is perfectly binding.
  \item if the probability is lower than some value  $\epsilon$ we say that the commitment function is $\epsilon$-statistically binding.
   \item if the probability is negligible in $k$ and $\mathsf{Commit}$ is polynomially bounded in $k$, we say that the commitment function is computationally binding.
 \end{enumerate}

  \item Hiding Property: For any message $m$ the commitment $c$ generated by $\mathsf{Commit}(m) \rightarrow (c,d)$ does not leak any information concerning $m$.
  
Concretely, we consider the following success probability
$$\Pr [m \leftarrow \mathcal{A}(c)|((c,d) \leftarrow \mathsf{commit}(m) ) ]$$
 for which we consider several cases:  
  
 \begin{enumerate}
 \item if the probability is equal to $\frac{1}{|M|}$ we say that the commitment function is perfectly hiding.
 \item if the probability is lower than some value  $\epsilon$ we say that the commitment function is $\epsilon$-statistically hiding.
 \item if the probability is negligible in $k$ and $\mathsf{Commit}$ is polynomially bounded in $k$, we say that the commitment function is computationally hiding.
 \end{enumerate}
 \end{itemize}

\subsection{Random Oracles}
Random Oracles are ideal objects that implement a uniformly distributed random function from the set $X$ to the set $Y$,  $\mathsf{RO}:X \rightarrow Y$ such as that for $x \in X, \mathsf{RO}(x)=y \in Y$ where $y$ is a random value. If the same value $x$ is given to the Oracle then the same value $y$ will be outputted. Random Oracles are usually used to prove security in system (the system is then called \textit{secure in the random oracle model}).

\subsection{Digital Signatures}
A digital signature scheme consist of a message $mess$ in the message space $M$, a signature $sig$ in the signature space $S$, private key $K_{s}$,a public key $K_{p}$, a Generation algorithm $\mathsf{Gen}$ with some parameters $Param,k$; a signing algorithm $\mathsf{Sign}$ and a verification algorithm $\mathsf{Verif}$ that are used a follow:

\begin{itemize}
  \item Generation: the generation of the pair of keys is done by the probabilistic function $\mathsf{Gen}(Param,1^k)\rightarrow (K_{s},K_{p})$.
  \item Signing: the signature of the message is done by the probabilistic function $\mathsf{Sign}(mess,K_{s}) \rightarrow sig$.
  \item The verification of the signature returns $\mathsf{true}$ or $\mathsf{false}$ if $(mess,sig)$ is valid with respect to the key pair $(K_{s},K_{p})$ or not with the (generally) deterministic function $ \mathsf{Verif}$:
  $$ \mathsf{Verif}(mess,sig,K_{p} ) \rightarrow (\mathsf{true} \textrm{ or } \mathsf{false})$$
\end{itemize}

Digital signatures offers the \textit{correctness} propriety which guaranties that the $\mathsf{Verif}$ algorithm returns $\mathsf{true}$ for any the signature generated by $\mathsf{Sign}$ if the correct pair of keys has been used.

From this we can deduce some other proprieties:

\begin{itemize}
 \item 	Authentication: gives the ability to the Verifier to check if the pretended author of the message really is the author. It also guarantees to the Signer that no one without $K_{s}$ could sign the message.
 \item Integrity: gives the ability to the Verifier to check if the message has been tempered between the time it has been signed and verified.
 \item Non-repudiation: takes away the ability of the Signer to deny a genuine signature to a Verifier.
 \item Soundness: a valid signature cannot be proven $\mathsf{false}$ and an invalid signature cannot be proven $\mathsf{true}$.
\end{itemize}

Traditional digital signature schemes are \textit{universally verifiable} meaning that anyone can verify if a signature is correct or not by using the public key. With this type of signature the non-repudiation is implicitly given with the authentication and the integrity.

\subsubsection{Security of digital signatures}

When we talk about security, we generally need to distinguish between an attacker who performs a \textit{Known Message Attack}, where the attacker can have access to messages and valid signatures form an oracle, and a \textit{Chosen Message Attack}, where the attacker can decide of messages and get a valid signatures of the messages from the oracle.

A system that is \textit{Chosen} message resistant offers a greater security than one that is only \textit{Known} message resistant.

There are several different security levels to consider while talking about signatures:

\begin{itemize}
  \item A signature is \textit{total break resistant} if there is no way for an attacker to find the private key while only knowing the public key.
  \item A signature is \textit{universal forgery resistant} if there is no way for an attacker to find a valid signature for a \textit{random chosen} message by only knowing the public key.
  \item A signature is \textit{existential forgery resistant} if there is no way for an attacker to find a valid signature for a \textit{chosen message} by only knowing the public key.
\end{itemize}

\subsubsection{Interactive Proof}
During an interactive proof between the Prover and the Verifier, the Prover convinces the Verifier of the validity of a given statement with a proof based on a secret. An Interactive proof can have the following proprieties:

\begin{itemize}
  \item Zero-Knowledge: the protocol does not disclose any information about the secret while being performed.
  \item Completeness: when the statement is $\mathsf{true}$, the Verifier accepts the proof.
  \item Soundness: If the statement is $\mathsf{false}$, the Prover cannot convince the Verifier that the statement is $\mathsf{true}$.
  \item Non-Transferability: a third-party cannot prove a statement to the Verifier without knowing the Prover's secret.
\end{itemize}

\subsubsection{Undeniable Signatures}
Undeniable signature proposed by Chaum and van Antwerpen  \cite{chaum1989undeniable} offers the ability to the Signer to keep his signature private (and therefore not universally verifiable). This type of signature scheme incapacitates anyone to verify a signature and to associate it to the Signer without his collaboration (it is called an invisible signature since one needs the private key to be able to distinguish a valid from an invalid signature). The signature verification is done by an interaction between the Signer and the Verifier, where the Signer can choose if he wants to authenticate the message and has to prove to the Verifier if the message is genuine or not. Since a Signer could very easily deny a valid signature, the signature must be undeniable so that no valid signature may be denied (for that reason in the common language we call \textit{invisible signature} undeniable signature).

An Undeniable Signatures scheme consist of a message $mess$ in the message space $M$, a signature $sig$ in the signature space $S$, a private key $K_{s}$, a public key $K_{p}$, a Generation algorithm $\mathsf{Gen}$ with some parameters $Param,k$, a signing algorithm $\mathsf{Sign}$, a Confirmation interactive protocol $\mathsf{Conf}$ and a Denial interactive protocol $\mathsf{Den}$ that are used a follow:

\begin{itemize}
  \item Generation: the generation of the keys is done by the probabilistic function Gen: $\mathsf{Gen}(Param,1^k ) \rightarrow (K_{s},K_{p})$.
  \item Signing: the signature of the message is done by the signing probabilistic function Sign: $\mathsf{Sign}(mess,K_{s}) \rightarrow sig$.
  \item The Confirmation protocol of the signature returns $\mathsf{true}$ if $(mess,sig)$ is \textit{valid} with respect to the key pair $(K_{s},K_{p})$ else $\mathsf{false}$ with the function Conf: $\mathsf{Conf}(mess,sig,K_{p}) \rightarrow (\mathsf{true}$ or $\mathsf{false})$
  \item The Denial protocol of the signature returns $\mathsf{true}$ if $(mess,sig)$ is \textit{invalid} with respect to the key pair $(K_{s},K_{p})$ else $\mathsf{false}$ with the function Den:$\mathsf{Den}(mess,sig,K_{p}) \rightarrow (\mathsf{true}$ or $\mathsf{false})$
\end{itemize}

Undeniable Signatures guarantees the following properties:

\begin{itemize}
  \item Unforgeability: a valid signature can only been generated with the knowledge of the secret key.
  \item Soundness for confirmation: an invalid signature cannot be proven $\mathsf{true}$.
  \item Soundness for denial: a valid signature cannot be proven $\mathsf{false}$.
  \item Zero-Knowledge: the confirmation and the denial protocols do not disclose any information about the secret while being performed.
  \item Non-Transferability: a third-party cannot prove a statement to the Verifier without knowing the Prover's secret.
  \item Invisibility: a third-party cannot distinguish a valid signature from an invalid one without knowing the $K_{s}$.
\end{itemize}

\subsection{Diffie-Hellman}
The Diffie-Hellman protocol is a key agreement protocol based on the discrete logarithm problem \cite{hellman1976new} , that has the particularity to be done over an unsecured channel without the two parties (Alice and Bob) having previously exchanged information.

The agreement works as follows: Alice and Bob need two prime numbers \textit{g} and \textit{p} such that \textit{g} is a primitive route modulo \textit{p} (\textit{g} and \textit{p} are not secret and can be publicly available). Alice picks a random number \textit{a} and Bob a picks a random number \textit{b}. Alice then generates $A = g^a \mod p$ that is sent to Bob and Bob generates $B = g^b \mod p$ that is sent to Alice. To compute the secret key \textit{k} Alice computes $k = B^a \mod p = (g^b)^a \mod p$ and Bob computes  $k = A^b \mod p = (g^a)^b \mod p$.

A third person who would have listened to the agreement would only know \textit{p, g, A} and \textit{B} which would not be sufficient to find the secret key \textit{k}.

The Semi-static Diffie-Hellman protocol works exactly the same way as the DH protocol except that \textit{A} is given over a authenticated channel to Bob prior the key agreement. Therefore with the Semi-static protocol, it is possible to guarantee the authenticity of Alice.

\subsection{AES}
AES is a semetric cryptography standard based on the Rijndael cipher published by Rijmen and Daemen \cite{daemen1999aes}. It is block cipher based (128,192 or 256 bits) by using Substitution-permutation network on each block in 10,12 or 14 rounds (depending on the key size). Since 2001 it is an open Standard widely implemented.

\subsection{Modes of operation}
To avoid leaking information while encrypting with block cipher we have to use mode of operation. There exists several different mode of operation available (ECB, CBC, OFB, and CFB) that are more or less secure. When using ECB, someone without the knowledge of the secret key can still detect repeating blocks. The other modes use an \textit{Initialisation Vector} (IV) and chaining between blocks to increase the security and avoid this problem.

\chapter{MOVA Signature Scheme}

The MOVA signature Scheme has been developed by Jean Monnerat and Serge Vaudenay and presented at Asiacrypt 2004 \cite{monnerat2004generic}. MOVA is  an undeniable signature scheme that has the particularity to generate very small signatures while keeping a good security level.

\section{Preliminaries} 
We will first go through some proof protocol definitions that are used in the MOVA Scheme. The proofs and the detailed definitions can be found in \cite{monnerat3691short}.

\subsection{Proof Protocol for the GHID Problem (GHIproof)}
We need to define a protocol in which the Signer $S$ can prove to the Verifier $V$ that a set $R = {(x_1, y_1), \ldots , (x_s, y_s)}$ with $x_i \in X$ and $y_i \in Y$ \textit{interpolates} in the group homomorphism $H : X \rightarrow Y$ with $d$= $|Y|$. The protocol could be done $l$ times or in one batch of with $l$ values (as done underneath).
\begin{enumerate}
  \item $V$ randomly picks $r_i \in_u X$ and $a_{i,j} \in_u \mathbb{Z}_d$ for $i= 1, \ldots , l$ and $j= 1, \ldots , s$. He then calculates $u_i = dr_i + a_{i,1}x_1 + \ldots + a_{i,s}x_s$ and
$w_i = a_{i,1}y_1 + \ldots + a_{i,s}y_s$ for $i = 1, \ldots , l$. He then sends to $S$ $u_1, \ldots , u_l$.
  \item First, $S$ verifies  that for $ i = 1, \ldots ,l, H(x_i) = y_i$ (if the homomorphism is valid for all values else he aborts the protocol). Secondly, he calculates $\tilde{w}_i = H(u_i)$ for $i = 1, \ldots ,l$. Thirdly $S$ commits to the calculated values $\mathsf{Commit}(\tilde{w}_1, \ldots , \tilde{w}_l) \rightarrow (c, d)$. Finally, $S$ sends to $V$  the commitment $com$.
  \item $V$ replies with the chosen values of $r_i$ and $a_{i,j}$ for $i=1, \ldots , l$ and $j= 1, \ldots , s$ to $S$.
  \item First, $S$ verifies that the $u_i = dr_i + a_{i,1}x_1 + \ldots + a_{i,s}x_s$ for $i=1, \ldots , l$ (if the equation is valid else he aborts the protocol). Secondly, allows $V$  to open $com$ by sending the $\tilde{w}_i$ for $i = 1, \ldots , l$ and $de$.
  \item $V$ verifies that $\tilde{w}_i = w_i$  for $i = 1, \ldots , l$ and opens the commitment with $\mathsf{Open}(\tilde{w}_1, \ldots , \tilde{w}_l, c, d)$. If the value is $\mathsf{true}$ the proof is valid else the proof is rejected.
\end{enumerate}

\subsection{Proof Protocol for the co-GHID Problem (co-GHIproof)}
We need to define a protocol in which the Signer $S$ can prove to the Verifier $V$ that a set $T = ((\hat{x}_1, \tilde{y}_1) , \ldots, (\hat{x}_t, \tilde{y}_t))$ with $\hat{x}_i \in X$ and $ \tilde{y}_i \in Y$ \textit{does not interpolate} in the group homomorphism $H : X \rightarrow Y$. The protocol could be done $l$ times or in one batch of with $l$ values (as done underneath). For the decision of non-interpolation to be taken it is enough that only iteration is proven wrong. $d$ is the order of $Y$ with smallest prime factor $p$ and $R = ((x_1, y_1), \ldots , (x_s, y_s))$ with $x_i \in X$ and $y_i \in Y$ that \textit{interpolate} in the group homomorphism $H : X \rightarrow Y$.

\begin{enumerate}
  \item Firstly, $V$ randomly chooses $r_{i,k} \in_u X, a_{i,j,k} \in_u \mathbb{Z}_p, \lambda_i \in_u Z_p$ with $i=1, \ldots ,l, j=1, \ldots , s, k=1, \ldots , t$. Secondly, $V$ calculates for every $i$ and every $k$:
$u_{i,k}= dr_{i,k}+ \displaystyle\sum_{j=1}^s a_{i,j,k}x_j + \lambda_i \hat{x}_k$ and $w_{i,k}= \displaystyle\sum_{j=1}^s a_{i,j,k}y_j + \lambda_i \tilde{y}_k$. (We will consider that the set $u=(u_{1,1}, \ldots , u_{l,t})$ and the set $w=(w_{1,1}, \ldots , w_{l,t})$.) Finally, $V$ send $u$ and $w$ to $S$.
  \item Firstly, $S$ calculates for $k=1, \ldots, t$ $ H(\hat{x}_k)=\hat{y}_k$ and checks that for one or more $k$, $\hat{y}_k \neq \tilde{y}_k$ (if not the protocol is aboard). Secondly $S$ calculates $H(u_{i,k})=v_{i,k}$ for $i = 1, \ldots ,l$ , $k = 1, \ldots , t$. Since for at least one $k$, $ \hat{y}_k \neq \tilde{y}_k$ and since $w_{i,k} - v_{i,k} = \lambda_i(\tilde{y}_k - \hat{y}_k)$ because of H homomorphic propriety, it is possible for $S$ to reveal $\lambda = (\lambda_1, \ldots, \lambda_l)$. (If $S$ does not find all the $\lambda_i$ it means that $V$ is not honest so $S$  picks random values of $\lambda_i$). Finally $V$ calculates $\mathsf{Commit}(\lambda)\rightarrow(c,d)$ and sends the commitment to $V$.
  \item $V$ replies with all the $r_{i,k}$'s and $a_{i,j,k}$'s.
  \item $S$ verifies that the $u_{i,k}$'s and the $w_{i,k}$'s have been calculated correctly by recalculating them with all the known values $r_{i,k}$'s and $a_{i,j,k}$'s. If the values are correct he replies with $\lambda$ and $de$ else he aborts the protocol.
  \item $V$ verifies that $\mathsf{Open}(\lambda, c, d)\rightarrow \mathsf{true}$  and verifies that the $\lambda$ found by $S$ are the correct ones, then accepts the proof, else he refuses it.
\end{enumerate}

\section{MOVA}
To be able to use a signing scheme, we need to follow the following steps: generating the keys, signing the document, verifying the signature (acceptance and denial). Below we will develop all of them.

\subsection{Domain Parameters}
To qualify to some standards, there are some parameters $Param$ that have to be set: $L_{key}$ would be the key length, $L_{sig}$ would be the signature length, $I_{con}$ would be the number iteration to confirm a valid signature and $I_{den}$ would be the number of iterations to deny a signature. We first choose two groups $X_{group}$  and $Y_{group}$  (with $d$ as the cardinality of $Y_{group}$) with a homomorphism $h$ function between them (the two groups and the homomorphic function can be completely defined by their parameters: $X_p,Y_p,H_p$); we also need 2 pseudorandom number Generators $\mathsf{PRNG_k}$ and $\mathsf{PRNG_s}$ function (that are modelled by random oracles and defined by their seed) that takes an element from $M$ (such as a seed $k \in M$) then generates a given number of element in the group $X_{group}$.

\subsection{The keys generation scheme}
The key generation is an operation that only has to be done once by the Signer from which he will produce a private key ($K_{s}$) that he will keep for himself and a public key ($K_{p}$) that he will be able to widely distribute.

\begin{enumerate}
  \item $S$ selects a $X_{group}$ and a $Y_{group}$ with an group homomorphism $h:X_{group} \rightarrow Y_{group}$ and computes the order $d$ of $Y_{group}$.
  \item $S$ randomly chooses a seed $k$ and calculates $\mathsf{PRNG_k}(k) \rightarrow (x_1, \ldots ,x_{L_{key}})=X_{gen}$.
  \item $S$ calculates $h(X_{gen})=(y_1, \ldots ,y_{L_{key}})=Y_{gen}$.
\end{enumerate}
$X_{gen}$ and $Y_{gen}$ are therefore subgroups of $X_{group}$ and $Y_{group}$.
\begin{center}
Our public key $K_{p}$ is then $(X_P,Y_P,d,k,Param,Y_{gen})$.

Our private key $k_{s}$ is then $(H_p)$.
\end{center}
To guarantee security, $L_{sig}$ must be long enough so that the probability of having an other homomorphic function that would map $X_{gen}$ to $Y_{gen}$ is as low as possible.
In an other variant, the choice of $k$ is done by a RA (Registration Authority) that would guarantee that a good value is chosen (the RA has also to check that the Signer is not trying to make to many registrations attempts, the RA would also needs to sign $k_p$ as a guarantee of authenticity and the signature of $k_p$ will be part of the public key).

\subsection{The signing scheme}
To be able to sign a message we first have to map our textual message $message$ to $mess \in M$ with a deterministic function $\mathsf{Map}: ASCII \rightarrow M$. We will need a pseudo-random generator $\mathsf{PRNG_s}:M \rightarrow X_{group}$ and a homomorphic function $h:X_{group} \rightarrow Y_{group}$.

\begin{enumerate}
  \item $\mathsf{Map}(message) \rightarrow mess$.
  \item $\mathsf{PRNG_s}(mess) \rightarrow (x_{mess_1},\ldots,x_{mess_{L_{sig}}})=X_{mess}$.
  \item $h(X_{mess})=(y_{mess_1},\ldots,y_{mess_{L_{sig}}})=Y_{mess}$
\end{enumerate}

\begin{center}
Then signed message is $(message,Y_{mess})$.
\end{center}
The signature is therefore $L_{sig}*log_2d$ bits long.

\subsection{Verification (confirmation)}
In this verification phase, the Signer $S$ proves to $V$ the that the signature $Y_{mess}$ is \textit{valid} for a message $message$ and a set of keys.

\begin{enumerate}
  \item $S$ and $V$ compute $X_{gen}$ based on the $K_{p}$'s value $k$: $PRNG_k(k) \rightarrow (x_1, \ldots ,x_{L_{key}})=X_{gen}$.
  \item $S$ and $V$ then generate $\mathsf{Map}(message) \rightarrow mess$ and $PRNG_s(mess) \rightarrow (x_{mess_1} \ldots,x_{mess_{L_{sig}}})=X_{mess}$.
  \item The protocol finally uses the GHIproof with $R= ((X_{gen}, Y_{gen}) || (X_{mess}, Y_{mess}))$.
\end{enumerate}

\subsection{Verification (denial)}
In this verification phase, the Signer $S$ proves to $V$ that  the signature $Y_{mess}$ is \textit{invalid} for a message $message$ and a set of keys.

\begin{enumerate}
  \item $S$ and $V$ compute $X_{gen}$ base on the $K_{p}$'s value $k$: $PRNG_k(k) \rightarrow (x_1, \ldots ,x_{L_{key}})=X_{gen}$.
  \item $S$ and $V$ then generate $\mathsf{Map}(message) \rightarrow mess$ and $PRNG_s(mess) \rightarrow (x_{mess_1}, \ldots,x_{mess_{L_{sig}}})=X_{mess}$.
  \item The protocol finally uses the co-GHIproof with $R= (X_{gen}, Y_{gen})$ and $T=(X_{mess},Y_{mess})$.
\end{enumerate}

\chapter{MOVA Train Tickets}

This chapter refers to the actual implementation in Java of the MOVA protocol. The project offers a Server to generate keys and signatures, and also acts as a Prover, a Web application that requests signatures to the Server and a Terminal application that acts as a Verifier.

\section{Structure}

\subsection{Terminology}

The \textit{Message} is the core of the train ticket containing the information about the journey that will be performed. The \textit{Signature} is a valid MOVA Signature in respect to a given message and a key-set. The train \textit{Ticket} is a SMS containing the Message and the Signature. The \textit{Server} is the service that will sign the messages and then verify their authenticity. The \textit{Terminal} is the embedded device that the train controller will have with him and that can establish a connection with the Server to verify a signature. The \textit{WebClient} is a web-service that allows people to buy a train ticket that can establish a connection with the Server to get a message signed. The \textit{Client} is the person who will have the SMS ticket and would present it to the train controller. The \textit{Device} is either a WebClient or a Terminal.

\subsection{Usual scenario}

In a usual scenario, the customer will purchase his ticket on the MOVA's train ticket website, which is JPS based generated by WebClient. The WebClient will then connect to the Server in order to ask him to generate a valid signature for the given message (text of the ticket). The Client will finally receive the train ticket by SMS. 

In the train, the  controller will enter in the Terminal the ticket received by the Client and the Terminal will connect to the Server which will either confirm or deny the ticket.
\\

For this implementation we do not send any SMS to the client but for a real deployment we could use a SMS service provider such as TrueSenses (\url{www.truesenses.com}) to receive tickets request by SMS and then to send the Tickets to the client's cell phone by SMS.

\begin{figure}
\centering
\includegraphics[width=8cm]{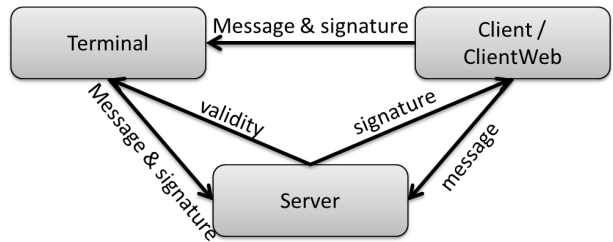} 
\caption{Direction of communication between the three parties}
\end{figure}

\section{Implementation}
The implementation is done in 4 different java projects. To use them, one should have JRE (\url{http://www.java.com/en/download/manual.jsp}) and a tomcat server (\url{http://tomcat.apache.org}) installed and running on the computer.

\begin{itemize}
\item \textbf{the Server Project} gives the Server service to the environment. It takes as argument the port number it has to listen to (if no port specified it will use the port 5000). Usually only one server should be launched per environment on a powerful computer so it can handle many simultaneous connections.
\item \textbf{The Terminal Project} gives a Terminal service to the environment. It takes as argument the IP address and port that it should use to connect to the Server (if not specified, it will use localhost and port 5000). The Terminal project should be launched on embedded devices with data connection. We can imagine that every train controller would have this kind of device with him.
\item \textbf{The ClientWeb Project} gives the WebClient service to the environment. At launch the user (it would be the administrator of the web site) must enter the IP address and the port to connect to the Server. The ClientWeb Project can be launched at any MOVA Train Ticket partner (such as Swiss or CFF) so they can offer the service to their clients.
\item \textbf{The MovaShared Project} is not a runnable project but contains all the shared classes of the three other ones. All the classes of MovaShared are already included in the other projects.
\end{itemize}
If the Terminal Project or the ClientWeb Project are launched before the Server project, they won't be able to connect and will have to be rerun to connect.

\subsection{Server}

\begin{figure}
\centering
\includegraphics[width=14cm]{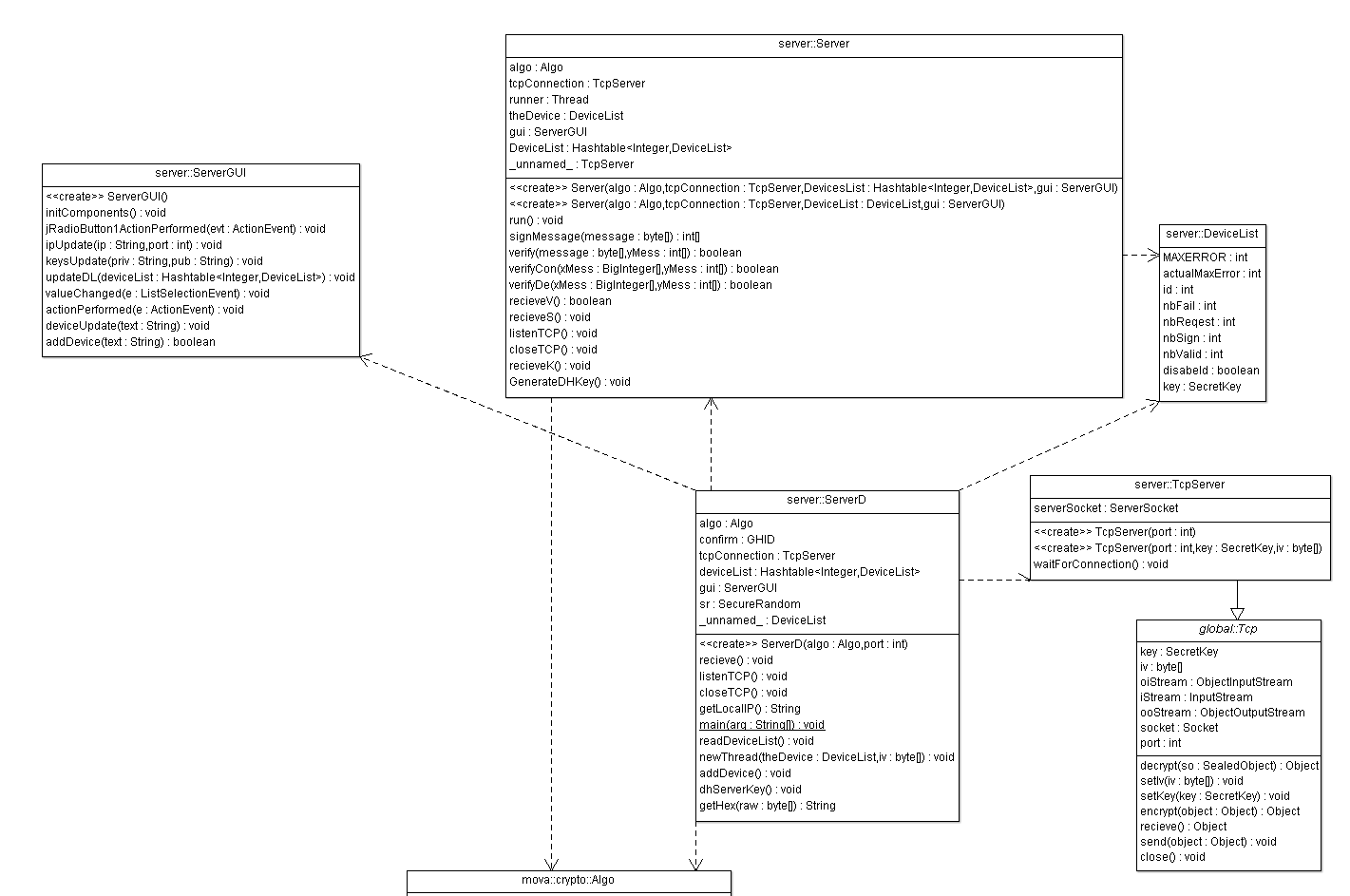} 
\caption{UML of the Server Package}
\end{figure}

At launch the Server will first generate its MOVA signing keys (or load the ones saved in the files \textit{keyPriv.key} and \textit{keyPub.key}) and load the list of all the known authorized devices (the list is in the file \textit{DeviceList.list} and points to other files with the complete information about specific devices). Then Server runs as a demon (\textit{ServerD.class}) always waiting for incoming connections on his listening port (by default 5000). When an incoming connection arrives, the Server verifies if the Device is in the DeviceList (with its unique ID): if the answer is positive, the Server replies with an open port number handled by \textit{Server.class} in a new thread; if the answer is negative, the Server requests from the Administrator an approval to add the new Device, creates a DeviceList entry, agrees on a AES secret key and sends back the unique ID of the Device.

Once connected to the new thread, the Server sends the IV for the AES encryption (after that, all the following communications are encrypted), then the Device can request from the Server to send his public key, sign a message or verify a message. When the interaction is finished, the thread connection is closed and the thread is killed.

At launch of the Server, there is a User Interface (\textit{ServerGUI.class}) window that shows the current Devices IDs with a short history of the interaction with all of them, a log of the actions of the Server and the ability to ban a specific Device (in case one of them is stolen or cheating).
When the Server starts for the first time, he generates the private and public key that are saved in the \textit{keyPriv.key} and \textit{KeyPub.key} files, so in case of crash or reboot from the Server the same key set can be used. If the Server for any reason wants to change his key set, the administrator must shut-down the Server, delete the .key files and relaunch the Server.

In order to to keep track of all the Devices that are authorised to communicate with the Server, the Server keeps a Hashtable of DeviceList (\textit{TerminalList.class}) that contains the unique ID of a Device, the number and type of requests made and its status. Every DeviceList's ID is saved into a the \textit{deviceList.list} file and points to other files containing the DeviceList data and is read at every launch of the Server. If the Server for any reason wants to remove the devices, the administrator must shut-down the Server, delete the \textit{deviceList.list} file and all the files with the \textit{dl} extension, then relaunch the Server.
To avoid too many attempts with false signatures by a Device, after a given number of failed tries (10 for demonstration purposes) the Device is banned and the administrator must rehabilitate it.

\begin{figure}
\centering
\includegraphics[width=8cm]{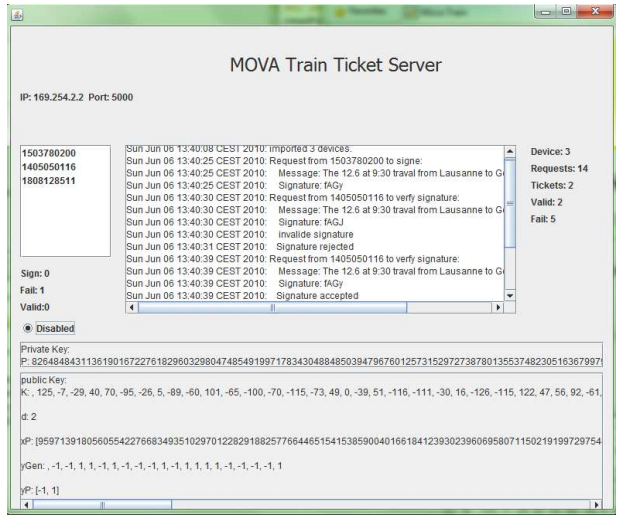} 
\caption{Screen shot of the Server's UI}
\end{figure}

\subsection{Terminal}

\begin{figure}
\centering
\includegraphics[width=14cm]{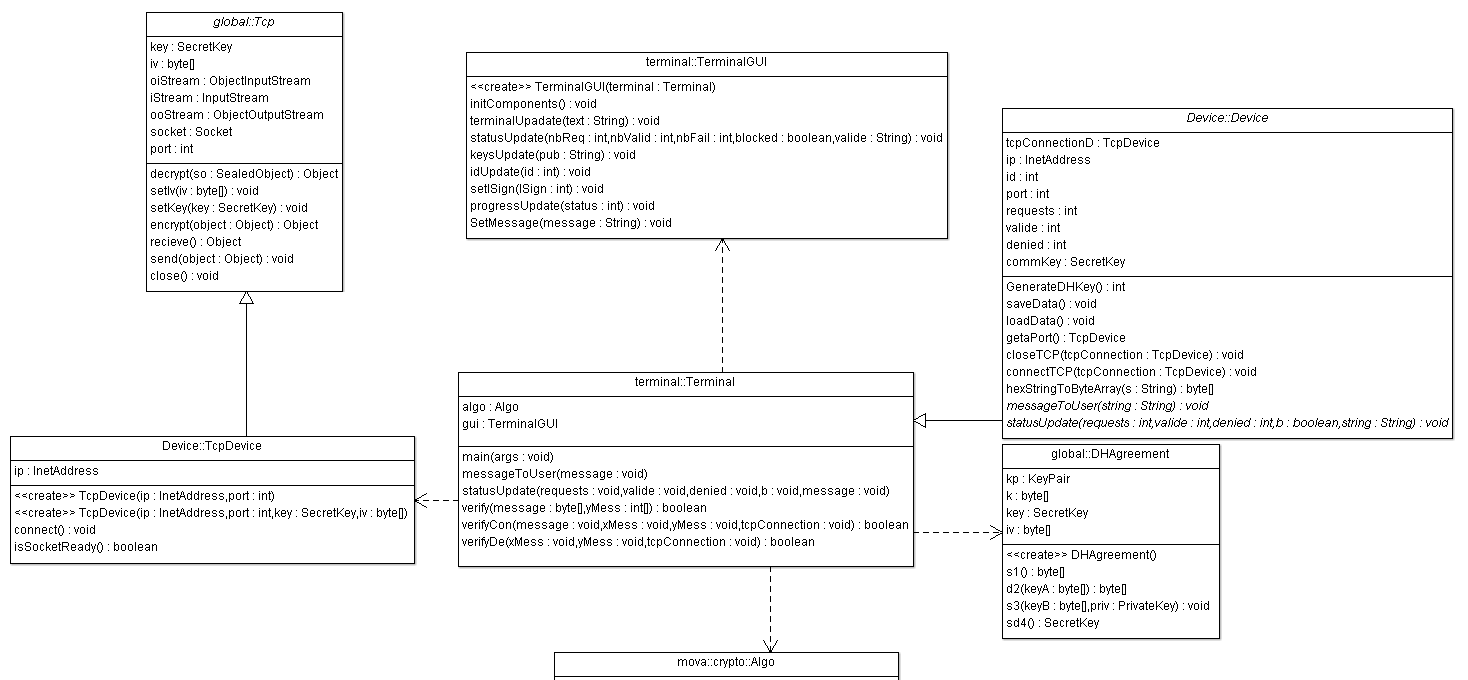} 
\caption{UML of the Terminal Package}
\end{figure}

The Terminal (\textit{Terminal.class}) can only do some very simple tasks, it is mainly here as an interface to the access the Server's services and keep some personal statistics. At launch it will request to the Server to receive an ID which will be used to authenticate at every interaction and they will agree with the DH protocol on a AES key. The user interface (\textit{TerminalGUI.class}) is intuitive and effective: it is constituted by two textboxes, one for the message and one for the signature, and a verify button which initiates the verification protocol. Once launched, the interface will display \textit{Valid} in green if the signature matches the message or display \textit{Invalid} in red if the message and the signature do not match.

\begin{figure}
\centering
\includegraphics[width=8cm]{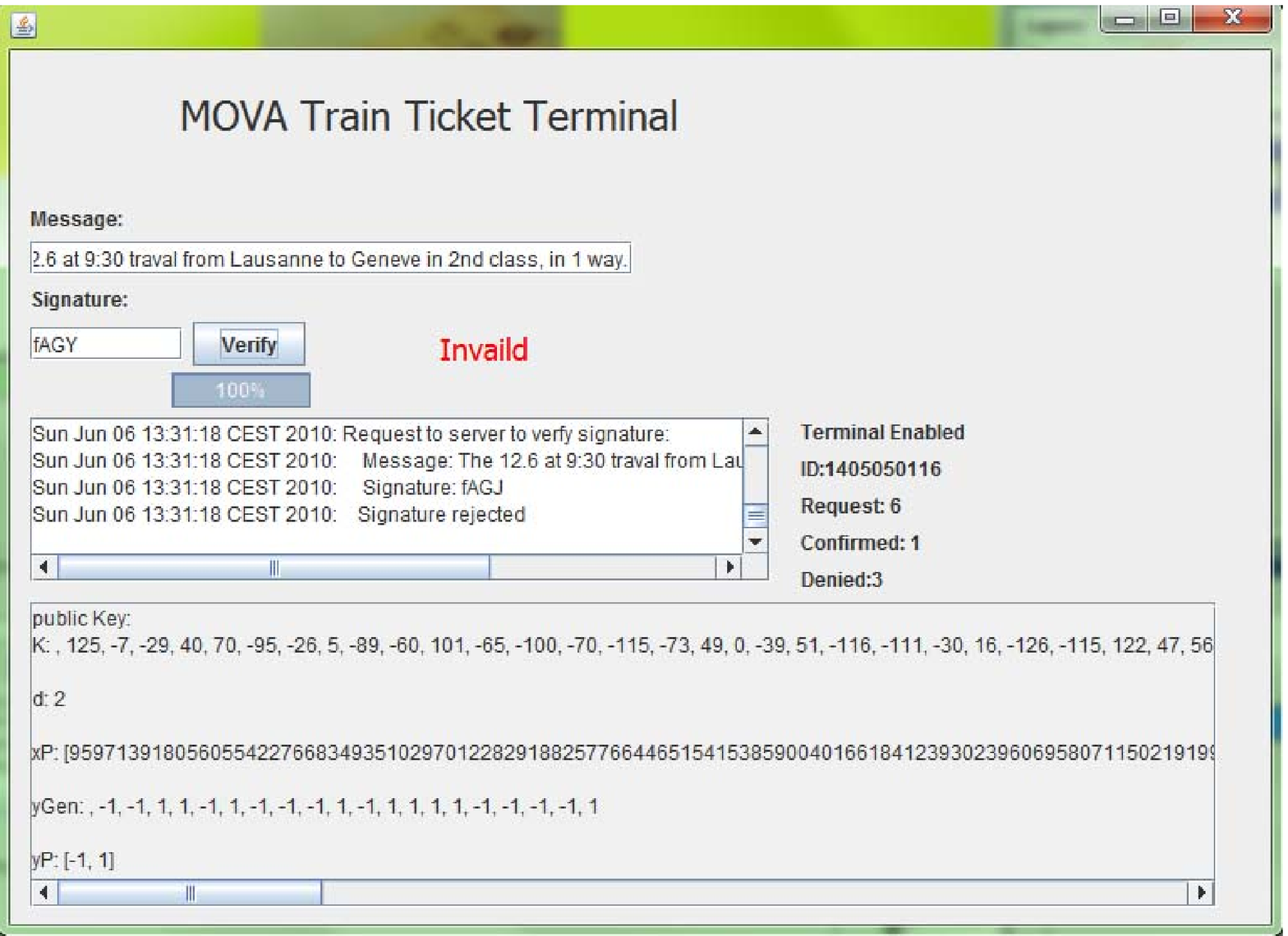} 
\caption{Screen shot of the terminal's UI}
\end{figure}

\begin{figure}
\centering
\includegraphics[width=14cm]{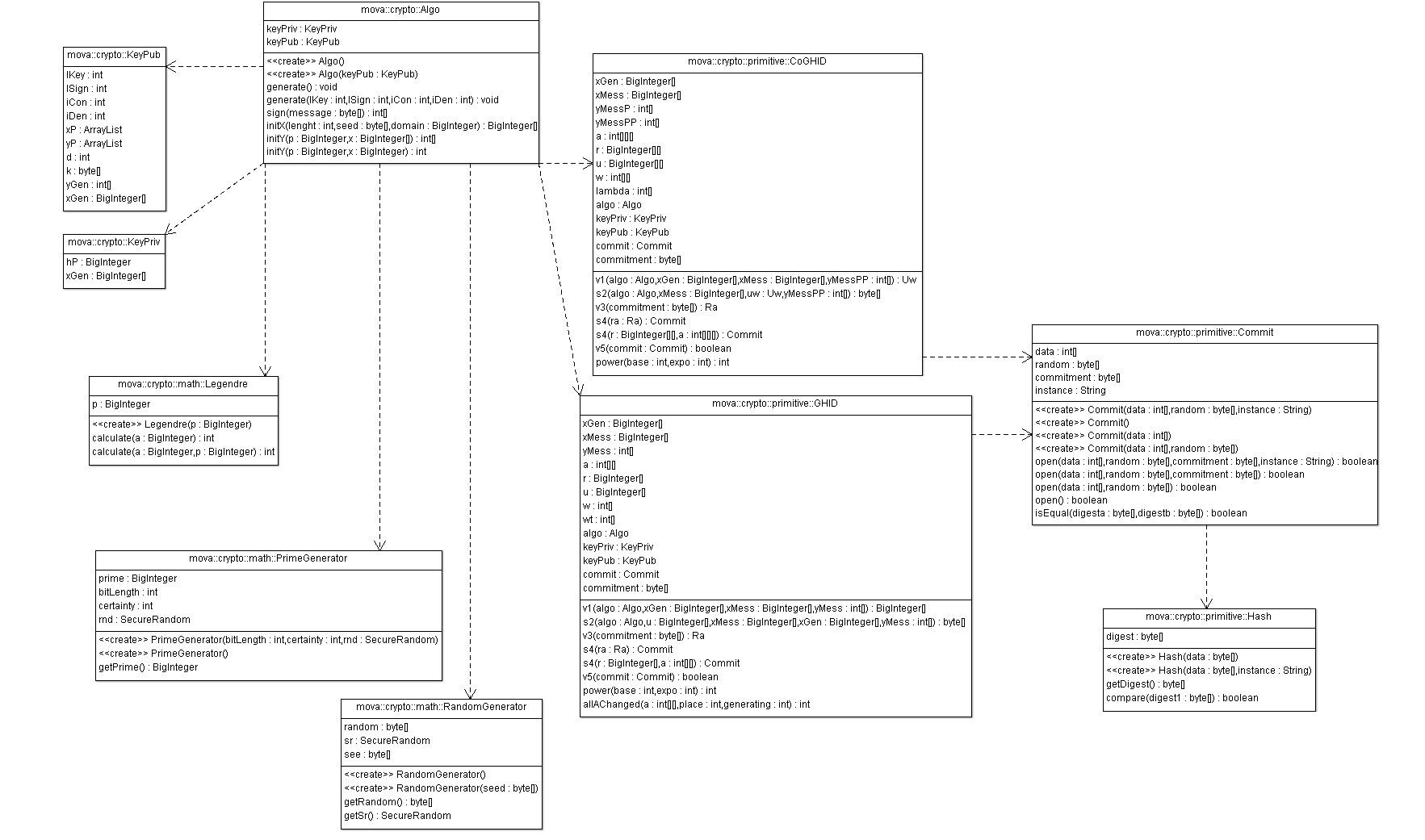} 
\caption{The UML of the MOVA algorithm used by Server and Terminal}
\end{figure}

\subsection{ClientWeb}

\begin{figure}
\centering
\includegraphics[width=14cm]{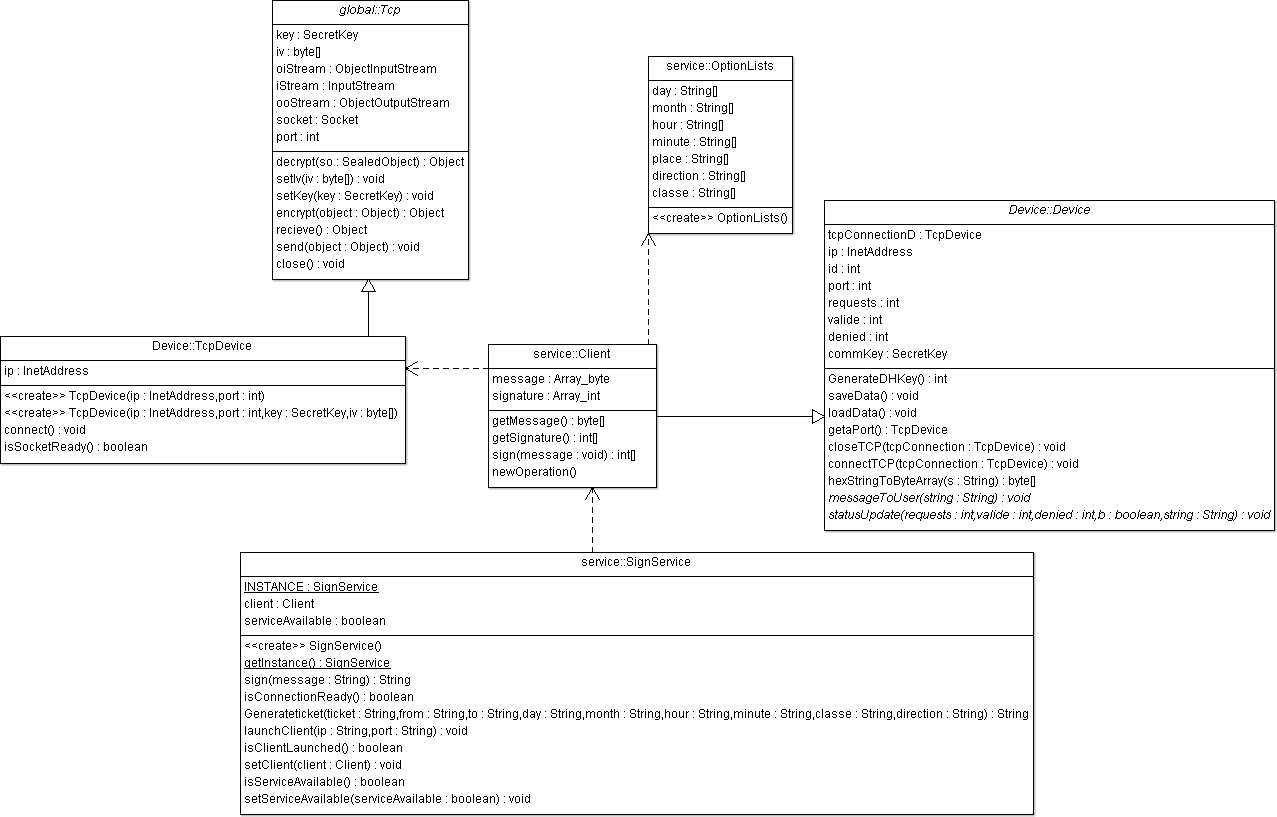} 
\caption{UML of the ClientWeb Package}
\end{figure}

The ClientWeb runs a singleton instance of the signing service (\textit{SignService.class}) that generates objects and text for the JSP pages, and creates an instance of the client service (\textit{Client.class}) that will handle the communication with the Server. At launch the client service will request from the Server to receive an ID which will be used to authenticate at every interaction and they will agree with the DH protocol on a AES key. The different lists from the drop down-menu (\textit{OptionList.class}) enable the User to generate the text from pre-existing information or the User (for demonstration purpose) can choose the text he wants (in a real deployment they should be a database behind the data for the drop-down list).

\begin{figure}
\centering
\includegraphics[width=8cm]{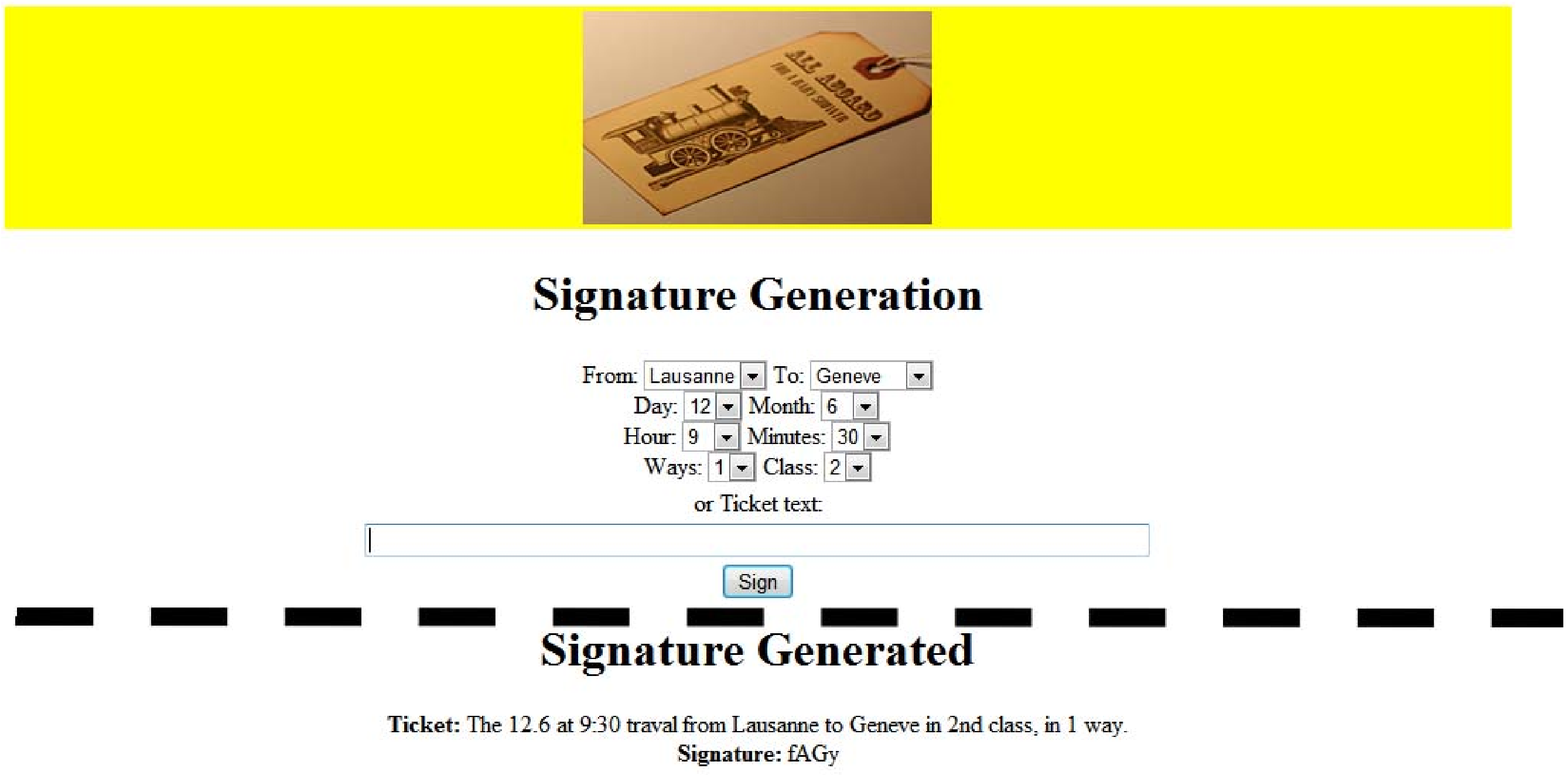} 
\caption{Screen shot of the ClientWeb's UI}
\end{figure}

\section{Usage}
To use the MOVA Train Ticket, one must first launch the Server, which will generate in the \textit{Server.jar} folder 4 files with the \textit{key} extension. The \textit{DhKpPub.key} is the DH public key, used for the Semi-static DH key-agreement. This file must be shared with the Terminal and the ClientWeb (it only has to be done the first time). To do so, one must insert it in the \textit{Terminal.jar} file and in the \textit{WEB-INF/classes} folder of the \textit{ClientWeb.war} file. (To edit Jar and War files easily one can use 7Zip \url{http://www.7-zip.org}). Then one has to deploy the \textit{ClientWeb.war} to the Tomcat Server, and launch the \textit{Terminal.jar} on a device.

\section{Communication Protocols}

\subsection{General}
All the communications are initiated by a Device (Terminal or ClientWeb). The Device sends his ID (-1 if he has no ID yet) in clear, then the Server replies with the AES IV and the port number on which the communication will then take place. The following communications all happen in a new Thread on the port number previously sent:

\subsection{Diffie-Hellman}
At the first launch of the Server, it generates its private $a$ and public $A=g^a$ mod $p$ keys and saves them to a file (\textit{dhKpPub.key}). (The Administrator must then put this file in all the devices.)
Since the Device has no ID, the \textit{GenerateDHKey()} method in the \textit{Server.class} and \textit{Device.class} are called.\

Client: first the Client loads the Server's public key (\textit{dhKpPub.key} and \textit{dhKpPriv.key}), secondly generates his own key-set, thirdly sends his public key to the Server, fourthly computes the secret key based on his private and the Server's public key, finally derives from the secret key an AES key and the IV.\

Server: first the Server loads his private key (\textit{dhKpPriv.key}), secondly receives the public key of the device, thirdly computes the secret key based on his private and the device's public key, finally derives from the secret key an AES key and the IV.\
The Client and the device have now a shared AES key, and the Server send the ID of the device (in an encrypted way).\

In all the future communications, the IV will be generated by a separated Secure random generator.

\begin{figure}
\centering
\includegraphics[width=8cm]{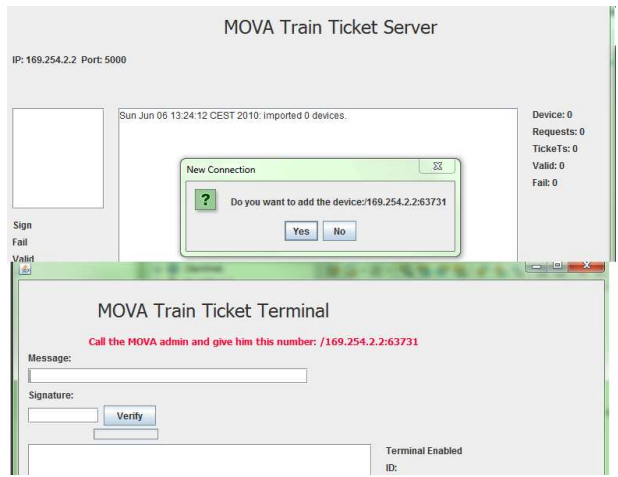} 
\caption{Request for approval to add a Device}
\end{figure}

\subsection{Public Key}
This protocol is only between the Terminal and the Server. All the communications are encrypted.
The Terminal calls the method \textit{getkey()} then sends the \textit{k} command to the Server. The Server then replies with his public MOVA key that is then displayed on the Terminal's UI.

\subsection{Sign}
This protocol is only between the ClientWeb and the Server. All the communications are encrypted.
The ClientWeb calls the method \textit{Client.sign(byte[] message)} with the message as parameter, then sends the \textit{s} command followed by the message. The Server then replies with the signature (in boolean) to the ClientWeb (the signature will been shown to the user in ASCII caracter).

\subsection{Verify}
The protocol is only between the Terminal and the Server. All the communications are encrypted.
The Terminal calls the method \textit{verify(byte[] message, String yMess)} with the message and the signature as parameter, then sends the \textit{v} command, the message and the the signature (converted to boolean) to the Server; then the Confirm protocol (GHID) starts. If at any stage of the protocol an error occurs, the message null is sent, and then the deny protocol (co-GHID) starts, otherwise the protocol finishes correctly.

\subsubsection{Confirm}
The confirm protocol works as described in the previous chapter. If at any stage the message received or the result return are not as expected,  the deny protocol starts.
If the protocol successfully finishes a \textit{Valid} massage appears on the Terminal and in the log of the Server. (The $w_i$ values chosen by the Terminal and found by the Server are printed in the java terminal).

\begin{figure}
\centering
\includegraphics[width=8cm]{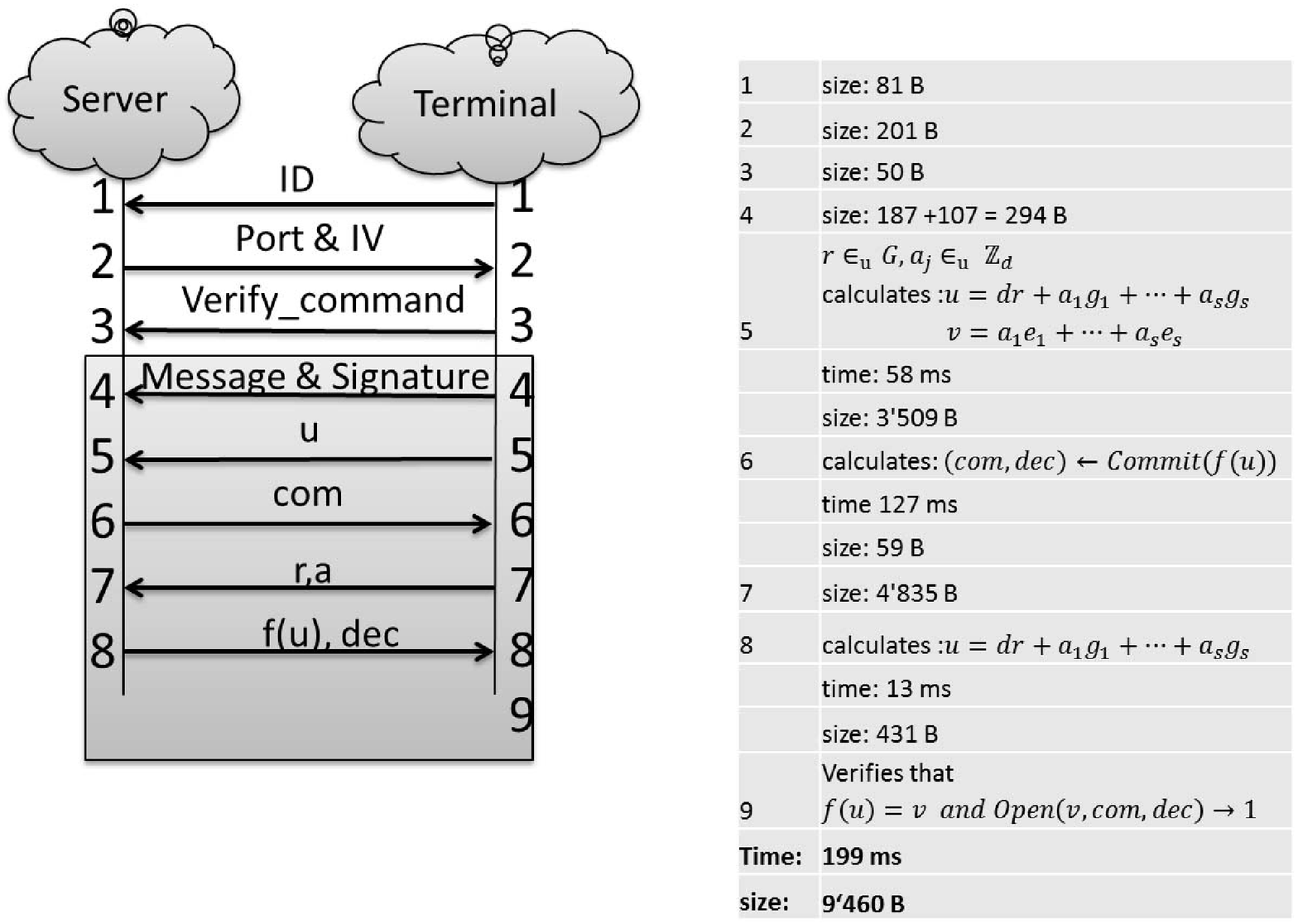} 
\caption{GHID protocol}
\end{figure}

\subsubsection{Deny}
The deny protocol works as described in the previous chapter. If at any stage the message received or the result return are not as expected, the machine is marked as cheating and is banned by the Server.
If the protocol successfully finishes a \textit{Invalid} message appears on the Terminal and in the log of the Server. (The $\lambda_i$ values chosen by the Terminal and found by the Server are printed in the java terminal).

\begin{figure}
\centering
\includegraphics[width=10cm]{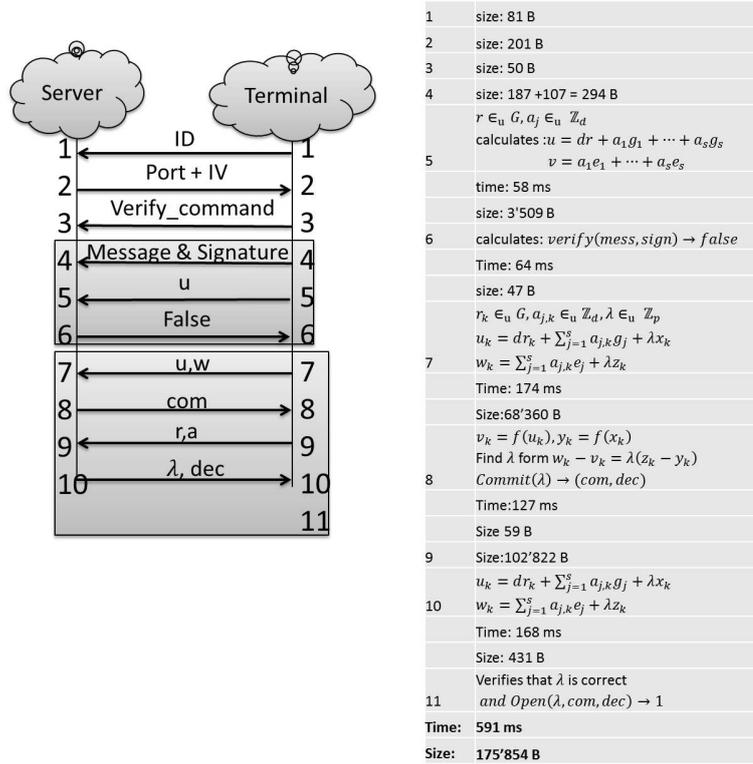} 
\caption{co-GHID protocol}
\end{figure}

\section{Implementation}
In this section we are going to see more into details how the methods have been implemented and which settings have been used.

\subsection{MOVA}

\subsubsection{Homomorphism function}
In the MOVA original paper several homomorphism functions have been proposed such as character on $\mathbb{Z}_{n}^{*}$ or RSA, each giving different proprieties and efficiencies.
In this project the priority has been put on efficiency by using the Legendre symbol $ \bigg( \dfrac{a}{p} \bigg)$, (with $n=pq$ and $p$,$q$ prime) which is defined as follow:
\[
\forall a\in \mathbb{Z}_p^\star
 \bigg( \dfrac{a}{p} \bigg) = \left\{ 
\begin{array}{l l}
  1 & \quad \text{if $a$ is a quadratic residue modulo $p$}\\
  -1 & \quad \text{if $a$ is a quadratic nonresidue modulo $p$}\\
\end{array} \right.
\]

\subsubsection{The Xgroup and Ygroup}
Since we have chosen to use the Legendre symbol as homomorphism function the \textit{Ygroup} = \{-1,1\}.
In the internal use of the application and while sending messages, the \textit{Ygroup} is represented as a boolean mapping $1 \rightarrow \mathsf{true}$ and $-1 \rightarrow \mathsf{false}$. When the signature must be sent in an SMS, it is then converted into ASCII characters. To avoid confusing some characters (such as 'o', 'O', '0') have been removed, so the mapping will only be done by converting 5 bits into 32 different characters. (all the methods are in \textit{ConvertY.class}).\

The \textit{Xgroup} are all the elements in $\mathbb{Z}_{n}^{*}$. We will generate two uniformly random prime numbers of 512 bit ($p$ and $q$ that will be kept secret,) to then calculate $n$ ($=pq$). It is computationally hard to calculate the symbol $ \bigg( \dfrac{a}{p} \bigg)$ by only knowing $n$ since it implies having to find the factorization of $n$.

\subsubsection{Security parameters}
Since we only take 5 bits to generate a ASCII character, it is important to use a small signature. In the MOVA original paper there is proof that, given a 20 bit signature:
\begin{itemize}
\item A Prover must have in his possession the secret key in order to interactively validate or deny a signature with a probability larger than $2^{-20}$.
\item If the probability of forging a valid signature is larger than $2^{-20}$ then the forger must have in his possession the secret key.\
\item If the Advantage of a Distinguisher between a valid and invalid message is larger than 0 then the Distinguisher has the secret key.\
\end{itemize}

In this implementation, \textit{iCon} and \textit{iDen} are set to 20 which means that for a signature to be accepted or rejected it has to pass the GHID or coGHID protocol 20 times (this represents a probability of $2^{-20}$ to pass the protocol when it should not).

 \subsection{Hash}
The Hash function used in this project comes from \textit{Hash.class}. It is based on the Java security library (\textit{java.security.MessageDigest}), uses \textit{SHA-256} and for any given message, it outputs a 32 bytes digest.

\subsection{Commitment}
The commitment function in this project comes from \textit{Commit.class}. It takes as input a message, concatenates it with a 128 random bytes then does a hash function on it that outputs a digest of 32 bytes (256 bits).\

The \textit{binding property} is given by the collision resistance of our hash function. With the birthday paradox, an attacker needs to compute $2^{n/2}$ ($2^{128}$) hash operations to likely find a collision.\

The \textit{hiding propriety} is given by the first pre-image residence of our hash function. An attacker needs to compute $2^{n}$ ($2^{256}$) hash operations to likely find a valid pre-image.

\subsection{Legendre}
The Legendre symbol is calculated in the \textit{Legendre.class} file. The method to do the calculation is (\textit{calculate(BigInteger a, BigInteger p)}) and applies this formula:
$$x \leftarrow a^{\frac{p-1}{2}} \mod p $$

\[
 \bigg( \dfrac{a}{p} \bigg) = \left\{ 
\begin{array}{l l}
  1 & \quad \text{$x=1$}\\
  -1 & \quad \text{$x=-1$}\\
\end{array} \right.
\]

\subsection{PseudoRandom Generator}
The Pseudo-random generator and the pseudo-random data are handled in \textit{RandomGenerator.class} which instantiates a \textit{SecureRandom} object (part of package java.security package). It uses a deterministic function with a random seed ;it can either self-seed getting it from the OS (usually using the timing of I/O events) \cite{rfc1750} or use a seed given in parameter. This implementation uses the java's recommended function \textit{SHA1PRNG} which has a period of $2^{160}$.

\subsection{Diffi-Hellman}
The Diffi-Hellman implementation is done in \textit{DHAgreement.class} following the Semi-static method. The public key of the Server, generated by the method \textit{s1()} is saved into a file and installed in all the devices. The key used to make the agreement is 1024 bits long.

The methods \textit{s1() and s3(byte[], PrivateKey)} are called by the Server and the method \textit{d2()} is called by the devices.

The implementation is inspired by \cite{oaks2001java} proposal.

\subsection{AES}
The AES implementation is done in \textit{DHAgreement.class} after the secret key has been generated. The secret key is used as a seed for a \textit{SecureRandom} that will generate the AES key (of 128 bits for portability reasons). The messages are then encrypted /decrypted in the \textit{Tcp.class} using the AES key generated in the DH agreement and and the IV set by the Server. The AES is used with CBC mode of operation and encrypts \textit{serialized object} into \textit{SealedObject}.

\chapter{Conclusion}
With this work, we have first seen some mathematical and cryptographic background to understand the MOVA scheme, then we have studied the MOVA scheme and identified its special properties that make this project possible. After all this being set, we have proposed a real implementation of the MOVA train Ticket service.

The Java implementation can prove the viability of such a system and shows that it can be done with an easy interface to generate, control and administrate train tickets and related devices.

There might still be some issues to real-world realisation that are not related to the MOVA security, such as assuring the unicity of a message and that passengers do not transmit tickets to each other; and the need of permanent data connectivity for the controller to always be able to verify tickets.

A future work could give deeper look into these problems.

\nocite{*}
\bibliographystyle{plain}
\bibliography{report}

\begin{thebibliography}{10}

\bibitem{rfc1750}
{Randomness Recommendations for Security, RfC-1750}, 1994.

\bibitem{chabaud1995links}
F.~Chabaud and S.~Vaudenay.
\newblock {Links between differential and linear cryptanalysis}.
\newblock In {\em Advances in Cryptology—EUROCRYPT'94}, page 356. Springer,
  1995.

\bibitem{chaum1989undeniable}
D.~Chaum and H.~van Antwerpen.
\newblock {Undeniable signatures, Advances in Cryptology-Crypto 1989, LNCS
  435}, 1989.

\bibitem{daemen1999aes}
J.~Daemen and V.~Rijmen.
\newblock {AES proposal: Rijndael}.
\newblock 1999.

\bibitem{goldreich2004foundations}
O.~Goldreich.
\newblock {\em {Foundations of cryptography: Basic applications}}.
\newblock Cambridge Univ Pr, 2004.

\bibitem{hellman1976new}
M.~Hellman.
\newblock {New directions in cryptography}.
\newblock {\em IEEE transactions on Information Theory}, 22(6):644--654, 1976.

\bibitem{knudsen1998java}
J.~Knudsen.
\newblock {\em {Java cryptography}}.
\newblock O'Reilly Media, Inc., 1998.

\bibitem{monnerat3691short}
J.~Monnerat.
\newblock {Short Undeniable Signatures: Design}.
\newblock {\em Analysis, and Applications. PhD thesis}, 3691, 2006.

\bibitem{monnerat2004generic}
J.~Monnerat and S.~Vaudenay.
\newblock {Generic homomorphic undeniable signatures}.
\newblock {\em Advances in Cryptology-ASIACRYPT 2004}, pages 1--6.

\bibitem{oaks2001java}
S.~Oaks.
\newblock {JAVA Security, Ed}, 2001.

\bibitem{vaudenay2005classical}
S.~Vaudenay.
\newblock {\em {A classical introduction to cryptography: applications for
  communications security}}.
\newblock Springer-Verlag New York Inc, 2005.

\end{thebibliography}

\end{document}